\documentclass[twocolumn,showpacs,preprintnumbers,amsmath,amssymb,superscriptaddress]{revtex4}
\usepackage{mathrsfs}
\usepackage{amssymb}
\usepackage{txfonts}
\usepackage{graphicx}
\usepackage{dcolumn}
\usepackage{bm}
\usepackage{subfigure}

\begin{document}


\title{Tunable ultranarrow linewidth of cavity induced by interacting dark resonances}

\author{Yandong Peng}
\affiliation{State Key Laboratory of High Field Laser Physics,
Shanghai Institute of Optics and Fine Mechanics, Chinese Academy of
Sciences, Shanghai 201800, China}%
\affiliation{Graduate University of Chinese Academy of Sciences,
Beijing 100049, China}
\author{Luling Jin}
\affiliation{State Key Laboratory of High Field Laser Physics,
Shanghai Institute of Optics and Fine Mechanics, Chinese Academy of
Sciences, Shanghai 201800, China}%
\affiliation{Graduate University of Chinese Academy of Sciences,
Beijing 100049, China}
\author{Yueping Niu}\thanks{Corresponding author.
E-mail: niuyp@mail.siom.ac.cn}%
\affiliation{State Key Laboratory of High Field Laser Physics,
Shanghai Institute of Optics and Fine Mechanics, Chinese Academy of
Sciences, Shanghai 201800, China}%
\author{Shangqing Gong}\thanks{Corresponding author.
E-mail: sqgong@mail.siom.ac.cn}%
\affiliation{State Key Laboratory of High Field Laser Physics,
Shanghai Institute of Optics and Fine Mechanics, Chinese Academy of
Sciences, Shanghai 201800, China}%

\date{\today}

\begin{abstract}

A scheme for getting a tunable ultranarrow linewidth of a cavity due
to an embedded four-level atomic medium with double-dark resonances
is proposed. It is shown that the steep dispersion induced by
double-dark resonances in the transparency window leads to the
ultranarrow transmission peak. Compared with the case of a
single-dark-resonance system, the linewidth can be narrowed even by
one order under proper conditions. Furthermore, the position of the
ultranarrow peak can be engineered by varying the intensity and
detuning of the control field.

\end{abstract}

\maketitle

Generally, narrower linewidth of cavity needs higher quality
resonator. While an electromagnetically induced transparency
(EIT)\cite{EIT} medium placed in an ordinary cavity can
significantly narrow the cavity linewidth\cite{In-EIT}. Though the
embedded medium spoils the optical quality factor of the resonator,
we can get a narrower linewidth with a less perfect cavity.  Xiao
and co-workers\cite{Xiao1} first experimentally demonstrated
cavity-linewidth narrowing by means of EIT with hot atomic-Rb vapor
in an optical ring cavity. After inserting atomic vapor, the finesse
of the cavity reduces from 100 to 51, but the cavity linewidth is
narrower by a factor 7 than empty-cavity linewidth. Later, in a cold
atomic-Rb system, Zhu $et\ al$.\cite{Zhu} observed a cavity
transmission spectrum with a central peak representing intracavity
dark state and two vacuum Rabi sidebands. The above experiments are
based on the single-dark EIT system. Steep normal dispersion and
almost vanishing absorption in the transparency window contribute to
the narrow peak in the transmission spectrum. In this case, the
position of the transmission peak locates at the resonance frequency
of the selected atomic medium, then loses manipulability. Many other
works have been carried out on the cavity+atom system, such as
optical bistability\cite{Agarwal} and multistability\cite{Xiao2},
quantum information and memory\cite{lukin00, Fleis, Duan},
photon-photon interaction\cite{PPI}, slow light\cite{slow, Hau}, and
so on.

As we all known, dark resonances are the basis of EIT and coherent
population trapping, etc. Double-dark resonance was first studied by
Lukin et al.\cite{lukin99}. The coherent interaction leads to the
splitting of dark states and the emergence of sharp spectral
features. Later, quantum interference phenomenon induced by
interacting dark resonances was observed in experiment\cite{2-exper}
 and different schemes of double-dark resonances were
explored\cite{Paspalakis, Pas06, Goren}. Some of us have done some
works based on double-dark resonances, such as high efficiency
4-wave mixing\cite{Niu1}, enhanced Kerr effect\cite{Niu2}, atom
localization\cite{Liu}, etc. In this paper, we find an efficient
scheme for the ultranarrow cavity transmission by interacting dark
resonances. In a tripod configuration, an additional transition to
the $\Lambda$-type EIT system by another control field causes the
occurrence of two distinct dark states. Interacting dark states
makes one of the transmission peaks much narrower. By proper tuning
of control fields, the ultranarrow spectrum can be one order of
magnitude narrower than that of a single-dark system. Moreover, its
position can be manipulated as we expect by adjusting the intensity
and detuning of the control field.

\begin{figure}
  \includegraphics[width=6cm]{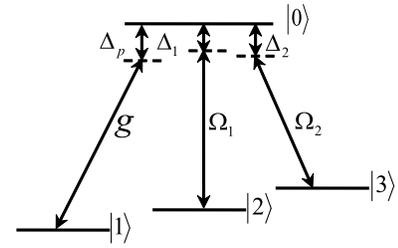}\\
   \caption{Schematic picture of the model configuration.
   Three lower states in the tripod system are coupled to the excited by three laser fields.}
\end{figure}

The atomic system under consideration is plotted in Fig. 1. Here,
$|0\rangle$ is an exited state and $|1\rangle$, $|2\rangle$,
$|3\rangle$ are three Zeeman ground sub-levels. A weak probe field
and two control fields with Rabi frquencies $g$, $\Omega_{1}$ and
$\Omega_{2}$ couple the transitions $|1\rangle-|0\rangle$,
$|2\rangle-|0\rangle$ and $|3\rangle-|0\rangle$ respectively.

The density matrix equations which describe the dynamics of the
system can be expressed as:
\begin{eqnarray}
\stackrel{.}{\rho }_{00}=- \left( {\gamma _{01}  + \gamma _{02} +
\gamma _{03} } \right)\rho _{00}  - ig \left( {\rho _{10} -
\rho _{01} } \right) \nonumber \\
\ \ \ \ \ \ \ \ \ -i\Omega _1 \left( {\rho _{20} - \rho _{02} }
\right) - i\Omega _2\left( {\rho _{30}  - \rho _{03} }
\right),\nonumber \\
\stackrel{.}{\rho }_{11}=\gamma _{01} \rho _{00} + \gamma _{21} \rho
_{22}  + \gamma _{31} \rho _{33}  - ig \left( {\rho _{01}  - \rho
_{10} } \right),\nonumber \\
\stackrel{.}{\rho }_{22}=\gamma _{02} \rho _{00}  - \gamma _{21}
\rho _{22}  + \gamma _{32} \rho _{33}  - i\Omega _1 \left( {\rho
_{02}  - \rho _{20} }\right),\nonumber\\
\stackrel{.}{\rho }_{33}=\gamma _{03} \rho _{00}  - \left( {\gamma
_{31}  + \gamma _{32} } \right)\rho _{33}
 -i \Omega _2 \left( {\rho_{03}  - \rho _{30} }\right),\nonumber\\
\stackrel{.}{\rho }_{10}=- \Gamma _{10} \rho _{10}  - ig \rho _{00}
+ ig \rho _{11} + i\Omega _1 \rho _{12}  + i\Omega _2 \rho _{13},\ \
\ \ \ \ \ \ \ \ \ \ \ \ \\
\stackrel{.}{\rho }_{20}=- \Gamma _{20} \rho _{20}  - i\Omega _1
\rho _{00}  + ig \rho _{21}  + i\Omega _1 \rho _{22}  +
i\Omega _2 \rho _{23},\nonumber\\
\stackrel{.}{\rho }_{30}=- \Gamma _{30} \rho _{30}  - i\Omega _2
\rho _{00}  + ig \rho _{31}  + i\Omega _1 \rho _{32}  + i\Omega _2
\rho _{33},\nonumber
\end{eqnarray}
\begin{eqnarray}
\stackrel{.}{\rho }_{12} = - \Gamma _{12} \rho _{12}  - ig
\rho _{02}  + i\Omega _1 \rho _{10},\nonumber\\
\stackrel{.}{\rho }_{13} = - \Gamma _{13} \rho _{13}  - ig
\rho _{03}  + i\Omega _2 \rho _{10},\nonumber\\
\stackrel{.}{\rho }_{23} = - \Gamma _{23} \rho _{23}  - ig \rho
_{03} + i\Omega _2 \rho _{20}.\nonumber
\end{eqnarray}
with $\rho _{kj}^ *=\rho _{jk}$ and the closure relation $
\sum\nolimits_j {\rho _{jj} }  = 1 \left( {j,k \in  \left\{
{0,1,2,3} \right\}} \right)$. Here ${\gamma _{0j} }$ represents the
spontaneous decay rate from state $\left| 0 \right\rangle$ to state
$\left| j \right\rangle$. $ \Gamma _{j0}  = \gamma _{0j}  - i\Delta
_j$, $ \Gamma _{12} = \gamma _{12} - i( {\Delta _p  - \Delta _1 }
)$, $ \Gamma _{13} = \gamma _{13}  - i( {\Delta _p  - \Delta _2 }
)$, $ \Gamma _{23}  = \gamma _{23} - i\left( {\Delta _1 - \Delta _2
} \right)$, and $ \Gamma _{kj}^ *=\Gamma _{jk} $, and $ \gamma _{jk}
\left( {j \ne k} \right)$ are the relaxation rates of the respective
coherences, where $ \Delta _p  = \nu _p  - \omega _{01} $, $ \Delta
_1  = \nu _1 - \omega _{02} $, and $ \Delta _2  = \nu _2  - \omega
_{03} $ denote the detunings of the corresponding fields.

Since a weak probe field couples the transition $ \left| 1
\right\rangle - \left| 0 \right\rangle$, its absorption and
dispersion properties are determined by the susceptibility of the
intracavity atomic medium. Considering initial condition (i.e.,
$\rho _{11}=1$), weak field approximation and ignoring the decay
between ground states (${\gamma _{12} }={\gamma _{13} }={\gamma
_{23} }=0$), the steady state linear susceptibility\cite{Paspalakis}
is given by
\begin{eqnarray}
\ \ \chi \left( {\Delta _p } \right) = -4\pi N\left| {\mu _{10} }
\right|^2 \frac{{ab\left( {\Delta _p ab - \Omega _1^2 b - \Omega
_2^2 a - i\gamma _{01}^2 ab} \right)}}{{\left( {\Delta _p ab -
\Omega _1^2 b - \Omega _2^2 a} \right)^2  + \gamma _{01}^2 a^2 b^2
}},\ \
\end{eqnarray}
where $N$ is the medium density, $a = \Delta _p  - \Delta _1$ and $
b = \Delta _p - \Delta _2 $. It is easy to see that when $a=0$ or
$b=0$, the atomic susceptibility is zero, which means a transparency
window at $ \Delta _p = \Delta _1$ or $\Delta _p  =\Delta _2$.
Further, we notice that the additional control field to a
$\Lambda$-type system leads to additional interference effects and
splits the transparency window into two if $\Delta _1 \ne \Delta
_2$\cite{Paspalakis}. By close inspection of Eq. (2), we also find
that the dependence of these transparency windows on the detunings
of the control fields provide the feasibility of manipulating the
atomic response at different frequencies.

\begin{figure}
  \includegraphics[width=7.6cm]{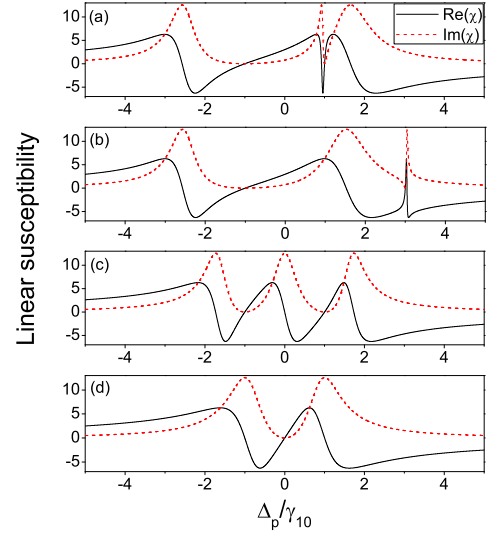}\\
   \caption{The linear susceptibility as a function of
   probe-field detuning ${\Delta _{p} }$.
   Re[$\chi$] (solid line) and Im[$\chi$] (dash line) for (a) $\Omega
   _1 =2.0$, $\Omega _2=0.3$, $\Delta _1 = -\Delta _2=-1.0$ and
    $\gamma _{12}=\gamma _{13}=\gamma _{23}=0.0001,
    $(b) as (a) but $\Delta _2=3.0$, (c) as (a) but $\Omega _1=\Omega
    _2 =2.0$, (d) as (a) but $\Omega _2=0.0$, $\Delta _1=\Delta _2=0.0.$}
\end{figure}
The linear susceptibility under different conditions are plotted in
Fig. 2 and some similar results have been reported in Ref.
\cite{Paspalakis}. Figures 2(a-c) depict the atomic susceptibility
for double-dark resonances and Fig. 2(d) for single-dark resonance.
For simplicity, all parameters are scaled by the decay rate $
{\gamma _{01} } $. When the control field $\Omega_2$ is tuned weaker
than the other field $\Omega_1$, the steep slope of the real ($\chi
'$) and image ($\chi ''$) parts of the atomic susceptibility emerge
in one narrow transparency window [see Fig. 2(a) or (b)]. Comparing
Fig. 2(a) or (b) with Fig. 2(d), we find that the dispersion
${{\partial {\chi '} } \mathord{\left/
 {\vphantom {{\partial  {\chi '} } {\partial \omega _p }}} \right.
 \kern-\nulldelimiterspace} {\partial \omega _p }}$
at the narrow transparency window is much larger than that in the
single-dark system. The additional coupling field makes the original
dark resonance split into two, then interacting dark resonances
leads to the dramatic change of the atomic susceptibility. What's
more, the position of the narrow transparency window depends on the
detuning of the weak control field. As we tune it from $\Delta _2=1$
to $\Delta _2=3$, the narrow transparency window moves from $\Delta
_p=1$ to $\Delta _p=3$ [see Fig. 2(a) and (b)]. Of course, when two
control fields are equal, the steep slope of the atomic
susceptibility tends to gentle, then we get a symmetrical
transparency window in Fig. 2(c).

Here, we consider the atomic medium within a vapor cell of length
$l$ in an optical ring cavity of length $L$, as shown in Fig. 3.
\begin{figure}
  \includegraphics[width=7cm]{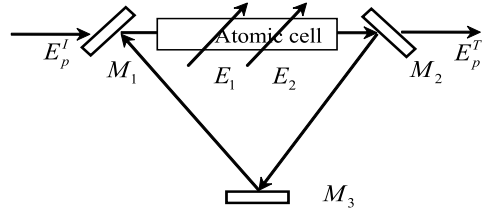}\\
   \caption{The ring cavity containing a atomic cell.
   $E_p^{I}$ and $E_p^{T}$ are the incident and the transmitted
   fields, respectively. $E_1$ and $E_2$ are two control fields.
   $M_1$, $M_2$ and $M_3$ are are reflecting mirrors.
}
\end{figure}
 The cavity field serves as the probe beam and the control
field is injected into the cavity with a polarization beam splitter
and copropagates through the vapor cell (not circulating in the
cavity) with the cavity field, which forms the two-photon
Doppler-free configuration\cite{Xiao3}. The cavity transmission can
be expressed as\cite{In-EIT}
\begin{eqnarray}
\ \ \ \ \ \ \ \ \ \ \ \ \ \ \ \ \ \ \  S\left( \nu  \right) =
\frac{{t^2 }}{{1 + r^2 \kappa ^2 - 2r\kappa \cos \left[ {\Phi \left(
\nu \right)} \right]}},\ \ \ \ \ \ \ \ \ \ \ \ \ \ \ \ \ \ \ \
\end{eqnarray}
where $t$ and $r$ are the transmissivity and the reflectivity of
both the input and the output mirrors $( {r=0.98, r^2  + t^2  = 1} )
$, and for simplify, we assume that mirror 3 has 100\% reflectivity.
$\Phi \left( \nu \right)={{\nu L} \mathord{\left/
 {\vphantom {{\nu L} c}} \right.
 \kern-\nulldelimiterspace} c} + {{\kappa l\chi '} \mathord{\left/
 {\vphantom {{\kappa l\chi '} 2}} \right.
 \kern-\nulldelimiterspace} 2}$ is the total phase shift, and $
\kappa  = \exp \left( { - \alpha l\chi ''} \right)$ is the medium
absorption per round trip.

It is instructive to examine the modified cavity linewidth. The rate
of the cavity linewidth\cite{Xiao1} $\Delta \omega _{d}$ of the
double-dark system to that of the single-dark system $\Delta \omega
_{s}$ reads
\begin{eqnarray}
\ \ \ \ \ \ \ \ \ \ \frac{{\Delta \omega _{d} }}{{\Delta \omega _{s}
}} = \frac{{1 + \omega _r ( {{l / {2L}}} )( {{{\partial \chi '_{s} }
/{\partial \omega _p }}} )}}{{1 + \omega _r ( {{l/{2L}}} )(
{{{\partial \chi '_{d} } /
  {\partial \omega _p }}} )}} \approx \frac{{{{\partial \chi '_{s} }/
  {\partial \omega _p }}}}{{{{\partial \chi '_{d} }/{\partial \omega _p
  }}}},\ \ \ \ \ \ \ \ \ \
 \end{eqnarray}
where $\omega _{r}$ is the resonant frequency of the cavity with
atomic system. The rate of $\Delta \omega _{d}$ to $\Delta \omega
_{s}$ approximatively equals the rate of the dispersion for
single-dark system to that for double-dark system. That is to say
that the cavity linewidth is reversely proportional to the
dispersion. When the intracavity atomic system is prepared in
double-dark states, its dispersion in the narrow transparent window
is much larger than that of single-dark system [see Fig. 2(a) and
(d)], then the cavity transmission peak can be much narrowed and the
ultranarrow linewidth of the cavity appears. Moreover, since the
detunings of the control fields determine the positions of the
transparency window, the position of the ultranarrow transmission
peak can be manipulated just by changing the detuning of the weak
control field.

\begin{figure}
  \includegraphics[width=8cm]{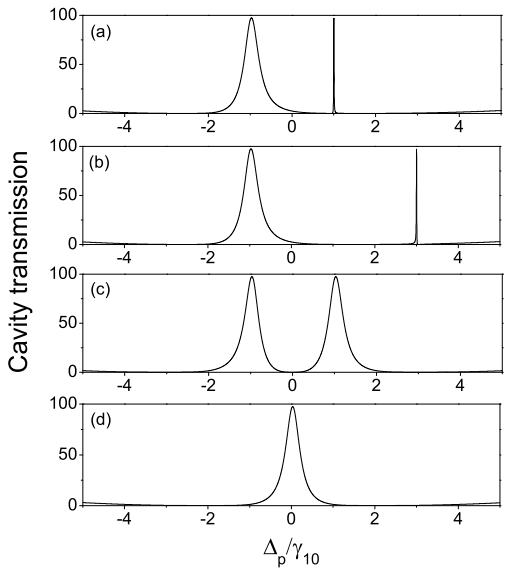}\\
   \caption{The cavity transmission spectrum as a function of
   probe-field detuning ${\Delta _{p} }$.  Ultranarrow spectrum
   for (a) $\Omega _1 =2.0$, $\Omega _2=0.3$, $\Delta _1 = -\Delta _2=-1.0$,
     and $r=0.98$, (b) as (a) but $\Delta _2=3.0$;
     (c) as (a) but $\Omega _1=\Omega _2=2.0$; (d) as (a) but $\Omega _2=0.0$, $\Delta _1=\Delta _2=0.0.$}
\end{figure}
In the following, we present the numerical simulation of the cavity
transmission spectrum in Fig. 4 by choosing the same parameters in
Fig. 2. First, let us see the ultranarrow spectrum induced by
double-dark resonances. When $\Omega_1=2$, $\Omega_2=0.3$ and $
\Delta_1=-\Delta_2=-1$, we get a sharp narrow transmission peak [see
Fig. 4(a)]. In this case the value of expression (4) is 0.083, which
means that the cavity linewidth is narrowed by one order.
Furthermore, with a different detuning of the weak control field,
the position of the ultrnarrow transmission peak is manipulated [see
Fig. 4(b)]. Second, when two control fields are equal, the right
peak is broadened because here the atomic susceptibility gets
gentle. Then we get a symmetrical double-peak transmission [see Fig.
4(c)]. At last, for comparison, we plot the single-peak transmission
of the cavity with the conventional single-dark atomic system in
Fig. 4(d). Obviously, its linewidth is much broader than that of the
ultranarrow spectrum. Physically, double-dark states exist in a
tripod system and the interacting dark resonances cause the steep
dispersion leading to the ultranarrow spectrum.

Besides, due to the system's symmetry, we can get the tunable
ultranarrow transmission peak on the left of the resonance frequency
by tuning the intensity and detuning of the other control field.
Then we can control the ultranarrow spectrum of the cavity for both
red and blue detunings, which means more manipulability at a broad
frequency range.

In conclusion, we proposed an efficient scheme for ultranarrow
cavity transmission spectrum induced by double-dark resonances. The
results show that the interacting double-dark resonances narrows the
transmission spectrum dramatically. Moreover, by adjusting the
intensity and detuning of the control field, the position of the
ultranarrow transmission peak can be manipulated. Close inspection
of the atomic susceptibility obviously shows that an additional
control field causes the splitting of the dark states and
interacting dark resonances lead to the steep dispersion which, at
last, results in the fine spectral feature. In experiment, the
tripod atomic system has already been used for beam
splitter\cite{Bergmannn} and sub-Doppler and sub-natural linewidth
absorption spectrum\cite{Gavra}, which means feasible to achieve the
tunable ultranarrow linewidth of cavity. This scheme may have
potential applications in high-resolution spectroscopy, laser
frequency stabilization and studying cavity-QED effects.

This work was supported by the National Natural Science Foundation
of China (Grant Nos. 10874194, 60708008), the Project of Academic
Leaders in Shanghai (Grant No. 07XD14030) and the Knowledge
Innovation Program of the Chinese Academy of Sciences.

\end{document}